\documentclass[twocolumn,showpacs,floats,floatfix,superscriptaddress,aps,pra]{revtex4-1}
\usepackage{amsfonts}
\usepackage{amssymb}
\usepackage{amsmath}
\usepackage{graphicx}
\usepackage{bm}
\usepackage{color}
\usepackage{epsfig}
\usepackage{calc}
\usepackage{ifthen}
\usepackage{subfigure}
\usepackage{graphicx}
\usepackage{epstopdf}
\usepackage{epsfig}
\usepackage{braket}
\usepackage{hyperref} 
\usepackage{hyperref}
\hypersetup{
    colorlinks = true,
    linkcolor = {blue},
    citecolor = {blue},
}

\begin{document}

\title{Shortcut to adiabatic light transfer in waveguide couplers with a sign flip in the phase mismatch} 
\date{\today } 

\begin{abstract}
Employing counterdiabatic shortcut to adiabaticity (STA), we design shorter and robust achromatic two- and three- waveguide couplers. We assume that the phase mismatch between the waveguides has a sign flip at maximum coupling, while the coupling between the waveguides has a smooth spatial shape. We show that the presented coupler operates as a complete achromatic optical switch for two coupled waveguides and as an equal superposition beam splitter for three coupled waveguides. An important feature of our devices is that they do not require larger coupling strength as compared to previous designs, which make them easier to realize in an experimental setting. Additionally, we show that the presented waveguide couplers operate at a shorter device length and are robust against variations in the coupling strength and the phase mismatch. 
\end{abstract}

\author{Wei Huang}
\affiliation{Engineering Product Development, Singapore University of Technology and Design, 8 Somapah Road, 487372 Singapore}

\author{L. K. Ang}
\affiliation{Engineering Product Development, Singapore University of Technology and Design, 8 Somapah Road, 487372 Singapore}

\author{Elica Kyoseva}
\affiliation{Institute of Solid State Physics, Bulgarian Academy of Sciences, 72 Tsarigradsko Chaussee, 1784 Sofia, Bulgaria}
\maketitle


\section{Introduction}

Waveguide directional couplers are important elements in integrated optics, with applications in power transfer between waveguides \cite{Syms1992}, mode conversion \cite{Sun2009}, polarization rotation \cite{Watts2005}, quantum communication \cite{Nikolopoulos2008} and many other practical fields \cite{Agrawal2007}. The electric field propagation in coupled waveguides can be accurately described within the coupled-mode theory (CMT) \cite{Huang1994} and recently, it was shown that the spatial dynamics of coupled waveguides is analogous to the temporal dynamics of quantum optical systems driven by external electromagnetic fields \cite{Yariv1990,Longhi2005}. Building on this analogy between quantum mechanics and wave optics, many optical systems to manipulate the propagation \cite{Longhi2009} and polarization \cite{Rangelov2015} of light were proposed based on quantum optical techniques. 

The most widely used quantum optical method in designing directional waveguide couplers is adiabatic evolution. Several adiabatic techniques such as stimulated Raman adiabatic passage (STIRAP) \cite{Vitanov2001} and rapid adiabatic passage (RAP) \cite{Bergmann1998,Huang2017} were used to achieve robust optical power switching between two, three and even an array of coupled waveguides \cite{Longhi2006,Longhi2007,Rangelov2012,Hristova2016}. The main advantage of these waveguide couplers is their robustness to parameter variations while maintaining the efficiency of power transfer at 100\%. However, adiabatic evolution requires slow change in the coupling parameters and overall dynamics, which necessitates impractically long devices length.

To speed up the adiabatic evolution, shortcut to adiabaticity (STA) was initially proposed in the context of quantum optical systems to produce the same final populations in a finite, shorter time \cite{Chen2010, Torrontegui2013, Chen2011}. More recently, STA techniques based on Lewis--Riesenfeld invariants were applied in wave optics to design shorter and robust coupled waveguides \cite{tseng2014, Ho2015, Tseng2014, Paul2015, Chung2015}. Another STA approach based on transitionless quantum driving (counterdiabatic driving) \cite{Berry2009, del2013} was also used to realize directional waveguides couplers \cite{Chen2016}, where an additional coupling at maximum coupling between the waveguides was required. 

In this paper, we apply counterdiabatic STA to the phase mismatch model with a sign flip at maximum coupling. The coupling of the phase mismatch model has a hyperbolic-secant shape while the phase mismatch is constant, with a sign flip at the coupling maximum. This model was previously used to design a two-waveguide coupler, which realizes complete achromatic all-optical switching \cite{Huang2014}. Here, we apply counterdiabatic STA to this model to design {\it a shorter and more robust directional coupler} for two and three waveguides that notably does not require an increase in coupling strength.

Our paper is organized as follows. In Sec. \ref{secII}, we review the coupled mode theory (CMT) and the phase mismatch model as applied to adiabatic evolution in coupled waveguides. In Sec. \ref{secIII} we apply STA to design a two-waveguide coupler and in Sec. \ref{secIV} we present numerical results about its performance. In the following Sec. \ref{secV} we extend the system to three coupled waveguides and use STA to propose a robust achromatic equal superposition beam splitter. Finally, we present our conclusions in Sec. \ref{secVI}.

\begin{figure} [tbhp]
\centering
\includegraphics[width=0.3\textwidth]{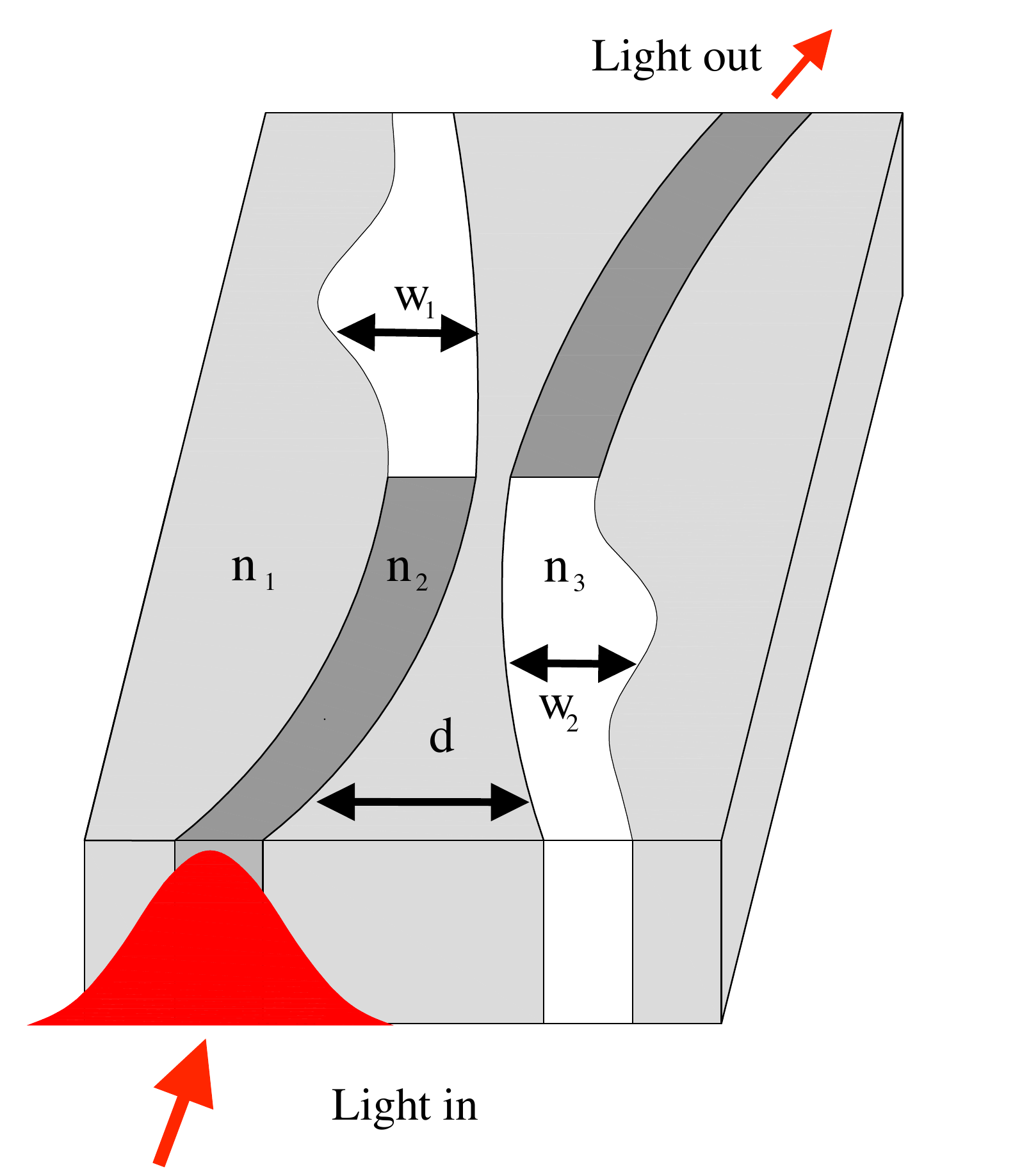}
\caption{(color online) A schematic of the waveguide structure with length $2L$ that is used for complete achromatic optical switching. Two evanescently coupled waveguides made of slabs with refractive indexes $n_2$ and $n_3$ are embedded in a medium with an index of refraction $n_1$. A Gaussian-shaped light beam is initially injected in the left waveguide and at the end the waveguide structure it is switched to the right waveguide.} 
\label{fig1}
\end{figure} 

\section{Adiabatic light transfer in coupled waveguides}
\label{secII}

We consider two evanescently coupled optical waveguides as shown in Fig. \ref{fig1}. Adopting the paraxial approximation, we describe the propagation of a monochromatic light beam in the waveguide structure of Fig. \ref{fig1} in the framework of the coupled-mode theory (CMT) \cite{Yariv1990}. The spatial propagation of the electric field amplitudes $c_1(z)$ and $c_2(z)$ in the $z$ direction is governed by a set of coupled differential equations, 
\begin{equation}
i\dfrac{d}{dz}\left[
\begin{array}{c}
c_{1}(z) \\
c_{2}(z)%
\end{array}%
\right] = \begin{bmatrix}
\Delta(z) & \Omega (z) \\
\Omega (z) & -\Delta(z)%
\end{bmatrix}%
\left[
\begin{array}{c}
c_{1}(z) \\
c_{2}(z)%
\end{array}%
\right] , 
\label{Schr}
\end{equation}
where $\Omega(z)$ is the coupling coefficient between the two waveguides and the phase mismatch $\Delta(z) = \big[\beta_1(z) - \beta_2(z) \big]/2$ is the difference between the corresponding propagation constants, $\beta_1(z)$ and $\beta_2(z)$. The absolute squares of the electric field amplitudes are the dimensionless light intensities in the waveguides, $I_{1,2} (z) = |c_{1,2} (z)|^2$, which are normalized to $I_1 (z) + I_2 (z) = 1$ in the lossless waveguides scenario. 

We consider the {\it phase mismatch coupling model}, 
\begin{eqnarray}
\Omega (z) &=&\Omega _{0}\,\text{sech}\left(2 \pi z/L\right) , \notag \\
\Delta (z) &=&\left\{
\begin{array}{c}
\Delta _{0} \\
-\Delta _{0}%
\end{array}%
\right.
\begin{array}{c}
(z<0) \\
(z>0)%
\end{array} , \label{parameters}
\end{eqnarray}
where $\Omega_{0}$ is the maximum coupling amplitude and $\Delta_{0}$ is a fixed phase mismatch. They are both positive and real. The total length of the waveguide structure is $2L$, while the middle point is at $z=0$. This coupling model is known to be analytically solvable and was previously used in Ref. \cite{Vitanov2007} to realize a complete population inversion in atomic systems. More recently, it was applied to achieve a complete light transfer between two waveguides and a beam splitter for three coupled waveguides \cite{Huang2014}. 

With the help of the unitary transformation $U_0$ (in the diabatic basis),
\begin{equation}
U_0=%
\begin{bmatrix}
\cos(\theta/2) & -\sin(\theta/2) \\
\sin(\theta/2)  & \cos(\theta/2)%
\end{bmatrix}, %
\end{equation}
we transform the diabatic basis $\mathbf{c}(z) = [ c_1(z), c_2(z)]^T$ to the adiabatic basis $\mathbf{a}(z) = [ a_1(z), a_2(z) ]^T$ according to $\mathbf{a} (z) = U_0^{-1} \mathbf{c}(z)$. The transformation $U_0$ is unitary and the mixing angle $\theta$ is defined as $\tan(\theta) = \Omega(z) / \Delta(z)$. The operator $H(z)$ in the adiabatic basis is given by $H_a(z) = U_0^{-1} H(z) U_0 - iU_0^{-1} \dot{U}_0$, where the overdot represents a derivative with respect to $z$. 

For the system evolution to follow the adiabatic path the adiabatic condition must be fulfilled. That is, the difference between the diagonal elements of $H_a (z)$ must be much larger than the off-diagonal elements. Parametrically, the adiabatic condition is satisfied when  
\begin{equation}
\dot{\theta } /2 \equiv \dfrac{ \dot{\Omega} \Delta - \dot{\Delta}\Omega} {2(\Omega^2 + \Delta^2)} \ll \sqrt{\Omega^2 + \Delta^2}, 
\end{equation}
which means that $\Delta(z)$ and $\Omega(z)$ must vary slowly with the spatial parameter $z$. We note that in relation to coupled waveguide devices the adiabatic condition entails long device lengths for the realization of high fidelity adiabatic light transfer.

\section{Shortcut to adiabatic light transfer in waveguides}
\label{secIII}

We use the shortcut to adiabaticity protocol to design optical switching devices with shorter characteristic lengths. The STA is achieved by introducing an additional coupling between the waveguides, described by $H_d (z)$, which is used to nullify the off-diagonal elements of the adiabatic operator $H_a (z)$. The additional coupling operator is $H_d (z) = i \sum_{j} \ket{\partial_{z} a_j} \bra{a_j} $, which in the basis $\mathbf{c}(z)$ is given by,
\begin{equation}
H_d(z)=%
\begin{bmatrix}
0 & -i \dot{\theta} /2  \\
i \dot{\theta} /2 & 0%
\end{bmatrix}.
\end{equation}
The total effective coupling operator is then $H_{\text{eff}}(z) = H(z) + H_d (z)$ with $H(z)$ being the coupling operator in the diabatic basis from Eq. \eqref{Schr}. We thus obtain
\begin{equation}
H_{\text{eff}}(z)=%
\begin{bmatrix}
\Delta(z) & \Omega(z) -i \Omega_a(z)  \\
\Omega(z) + i \Omega_a(z) & -\Delta(z) %
\end{bmatrix},%
\label{eq6}
\end{equation}
where the additional coupling term is $\Omega_a (z) \equiv \dot{\theta} /2$ and $\Omega(z)$ and $\Delta(z)$ are the coupling and phase mismatch of the phase mismatch model from Eq. \eqref{parameters}.

As the coupling from Eq. \eqref{eq6} is imaginary, which is not physical for a coupled waveguide system, we use the transformation matrix $U_1$,
\begin{equation}
U_\phi=%
\begin{bmatrix}
e^{-i \phi/2} & 0   \\
0 & e^{i \phi/2}  %
\end{bmatrix},
\end{equation}
with $ \tan(\phi)= \Omega_a (z)/ \Omega (z)$, to transform $H_{\text{eff}}(z)$ such as to remove the phase from the coupling terms. We thus obtain,
\begin{equation}
H_{\text{eff}}(z)=%
\begin{bmatrix}
\Delta_{\text{eff}}(z) & \Omega_{\text{eff}}(z)  \\
\Omega_{\text{eff}}(z) & -\Delta_{\text{eff}}(z) %
\end{bmatrix},%
\end{equation}
where 
\begin{eqnarray}
&& \Omega_{\text{eff}} (z) = \sqrt{\Omega(z)^2 + \Omega_a (z)^2}, \notag \\
&& \Delta_{\text{eff}}(z) =  \Delta(z) - \dot{\phi}(z)/2.
\label{shortcut}
\end{eqnarray}



If the additional coupling $\Omega_a(z)$ between the waveguides is strong enough, the effective Hamiltonian $H_{\text{eff}}(z)$ can in fact follow the adiabatic path in an arbitrary short time.  However, there is a physical limitation stating that the additional modifying coupling can not be larger than the original one, that is, $|\Omega_a (z)| \le |\Omega (z) | \le |\Omega_0 (z)|$ \cite{Chen2010}. To check if the STA model fulfills this inequality, we turn our attention to the behavior of the coupling parameter at the phase mismatch point, $z=0$. It is easy to see that $\lim_{z \rightarrow \pm 0} \Omega_a =0$ and we obtain that $\Omega_{\text{eff}} (0) = \Omega_0$. Therefore, unlike previous couplers based on counterdiabatic STA \cite{Paul2015}, the proposed coupler does not require an increase in the coupling strength at maximum coupling. 

\begin{figure} [tb]
\centering
\includegraphics[width=0.5\textwidth]{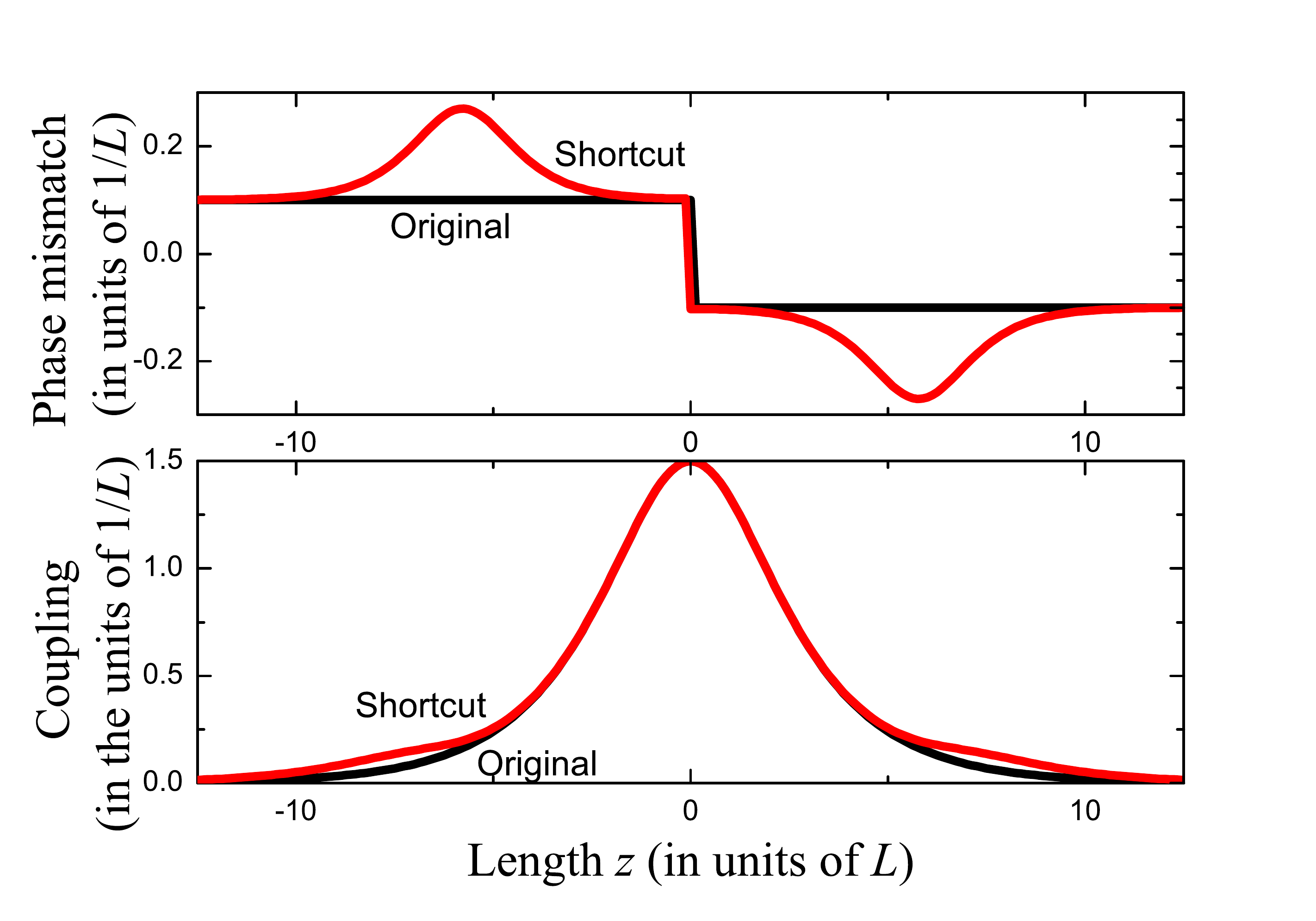}
\caption{Coupling strength and phase mismatch of both models as a function of the device length $z$. The original phase mismatch model parameters $\Omega(z)$ and $\Delta(z)$ from Eq. \eqref{parameters} are plotted with a black line, while the effective $\Omega_{\text{eff}}(z)$ and $\Delta_{\text{eff}}(z)$ from Eq. \eqref{shortcut} are plotted with a red line. We set $\Delta_0=0.1 \; \text{mm}^{-1}$ and $\Omega_0 = 1.5 \; \text{mm}^{-1}$, which corresponds to an input wavelength of 850 nm and the device length is 25 mm.}
\label{fig2}
\end{figure}

\begin{figure} [htbp]
\centering
\includegraphics[width=0.5\textwidth]{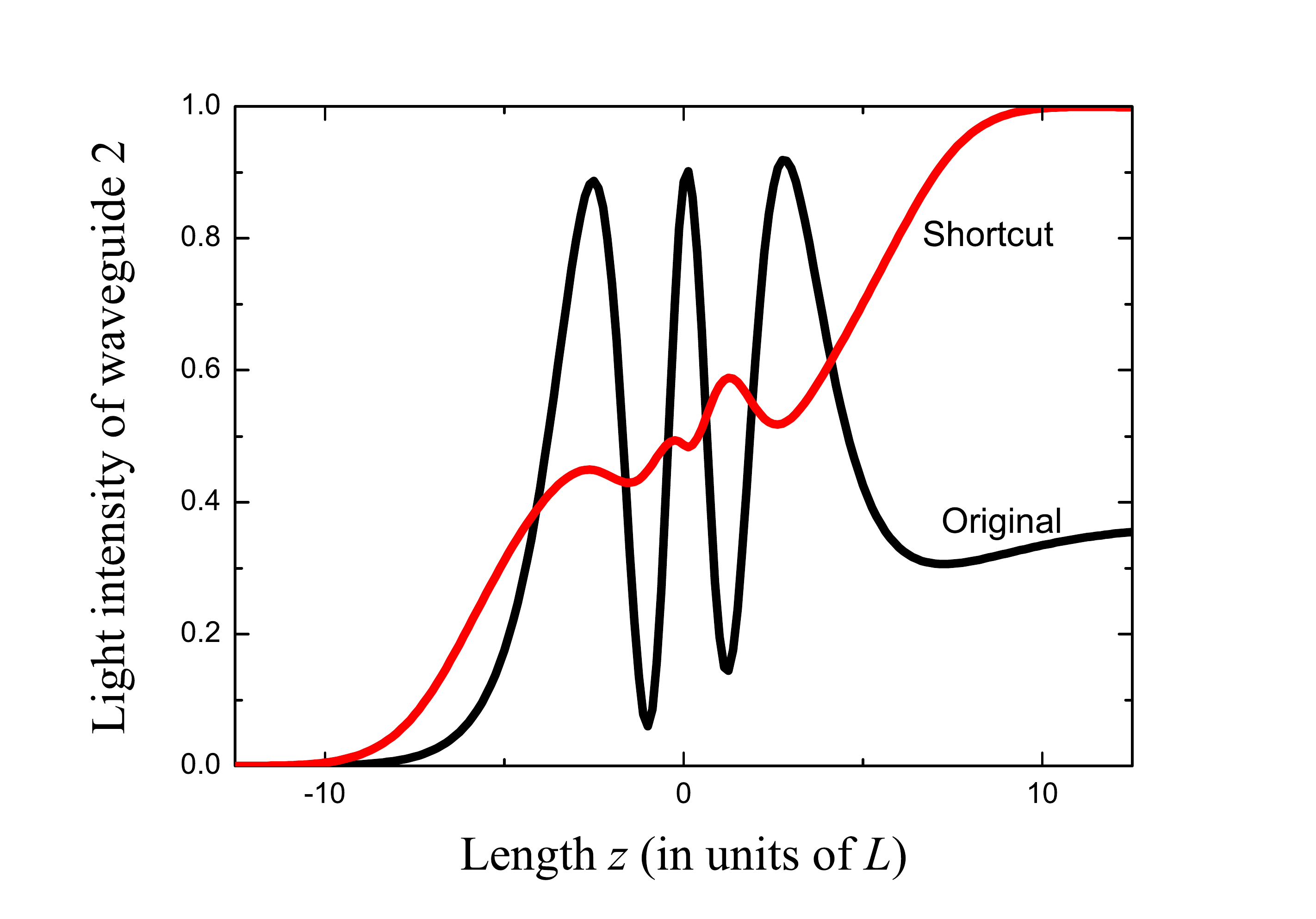}
\caption{The light intensity of waveguide 2, $I_2 (z) = |c_2 (z)|^2$, along the device length $z$ for the original (black line) and the shortcut to adiabaticity (red line) phase mismatch models with the coupling parameters from Fig. \ref{fig2}.}
\label{fig3}
\end{figure}

We plot the phase mismatch and the coupling as a function of the device length $z$ for the original phase mismatch model (Eq. \eqref{parameters}) and for the STA model (Eq. \eqref{shortcut}) and show them in Fig. \ref{fig2}. We set the device length at 25 mm and $\Delta_0 =0.1 \; \text{mm}^{-1}$, and $\Omega_0 = 1.5 \; \text{mm}^{-1}$, which corresponds to an input wavelength of 850 nm \cite{Ciret2012, Ciret2013}. Subsequently, for Fig. \ref{fig3} we assume that the input light is injected into the waveguide 1 and we plot the light intensity, $I_2 = |c_2 (z)|^2$, of waveguide 2. We assume the same waveguides parameters, as in Fig. \ref{fig2}. It is clear that the light transfer for the STA system is much more effective, as compared to the original one. 

The geometry of the waveguide coupler is determined by the coupling strength $\Omega _{\text{eff}}(z)$ and the phase mismatch $\Delta _{\text{eff}}(z)$ parameters. The separation distance $d$ between the two waveguides can be well fitted by the hyperbolic secant form of coupling strength $\Omega _{\text{eff}}(z)$. In addition, the connection between the phase mismatch $\Delta_{\text{eff}} (z)$ and the difference between the widths of the two waveguides, $\delta W = W_1 - W_2$, is given by a linear relation. The engineering of a sign flip in the phase mismatch at the maximum coupling point can be realized by switching the materials of the two waveguides. This would realize a swap of the propagation constants of the two waveguides, $\beta_{1,2} (z = -0) \rightarrow \beta_{2,1} (z = +0)$, and thus, the desired sign flip in $\Delta_{\text{eff}}(z)$.

\section{Performance and advantages}
\label{secIV}

\subsection{Shorter device length}

\begin{figure} [tbhp]
\centering
\includegraphics[width=0.5\textwidth]{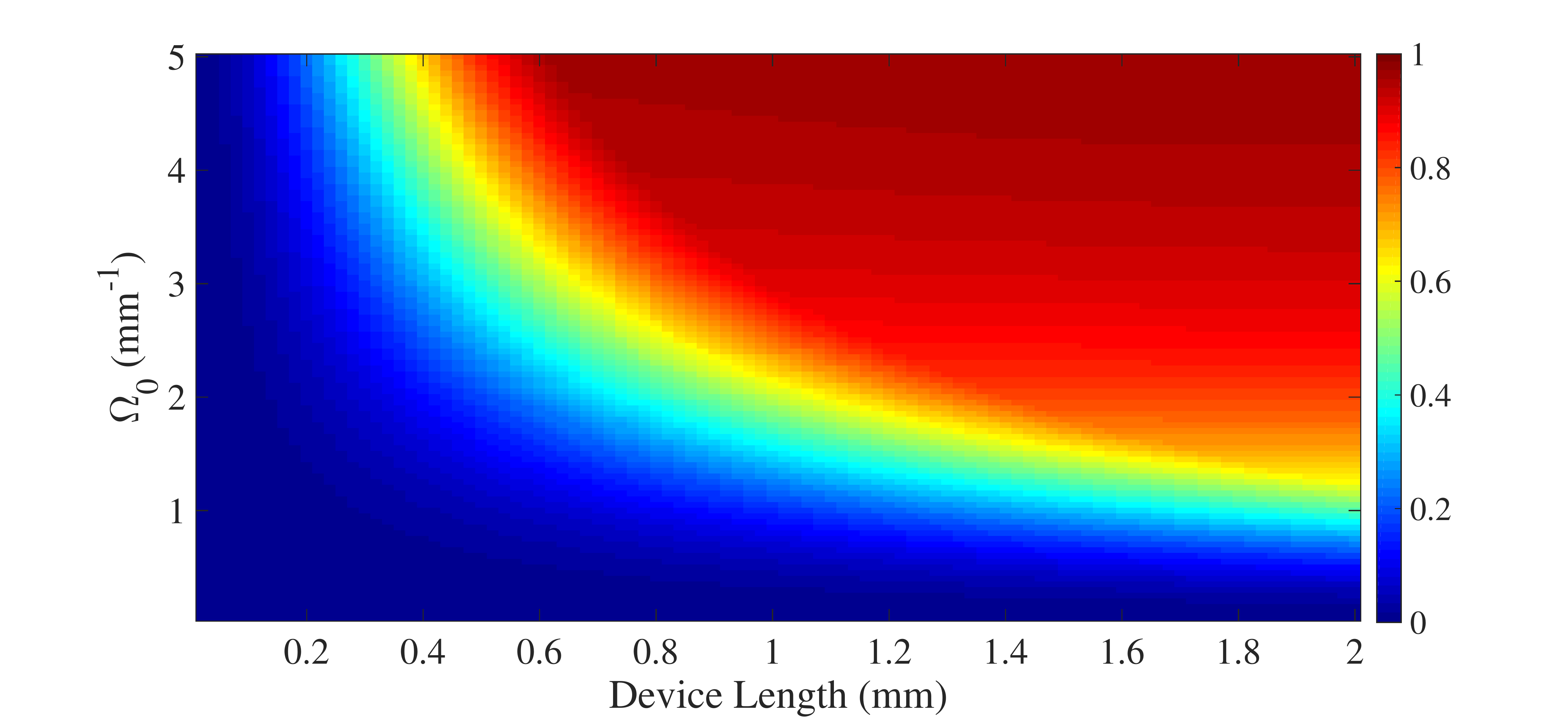}
\includegraphics[width=0.5\textwidth]{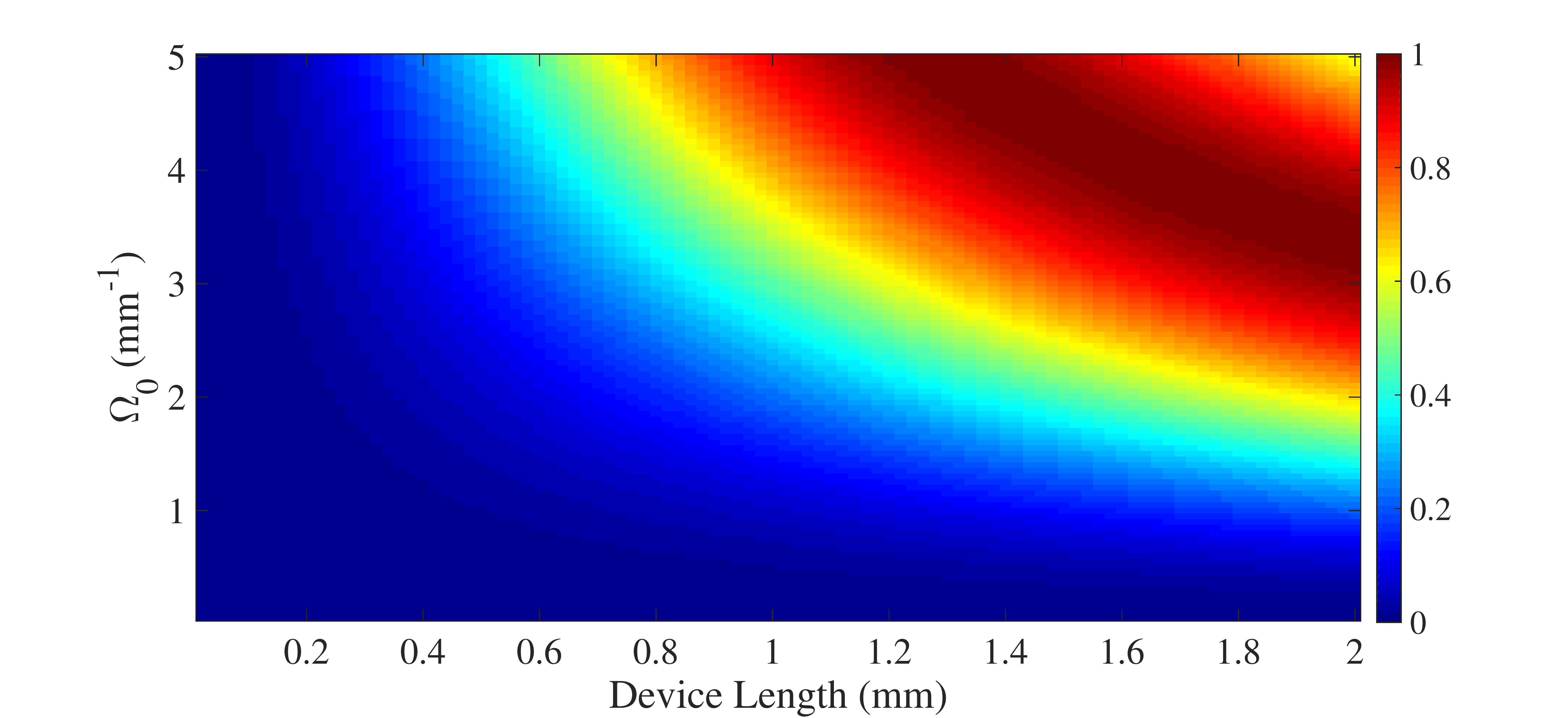}
\caption{(color online) Contour plots of the light intensity transfer at the end of the device, $I_2(L)$, for a waveguide coupler with a sign flip in the phase mismatch with STA (top frame) and without (bottom frame). The phase mismatch is set to $\Delta_0=1 \; \text{mm}^{-1}$, while we vary the maximum coupling strength $\Omega_0$ and the device length $2L$.}
\label{fig4}
\end{figure}

To compare the performance of the proposed STA waveguide coupler to the original one, we show the contour plots of the light intensity at the end of the device at waveguide 2, $I_2 (L)$, as a function of $\Omega_0$ and the device length $2L$. We assume that initially light was input in waveguide 1, $I_1 (-L) = 1$, and we solve the coupled differential equations from Eq. \eqref{Schr} numerically. The results are presented in Fig. \ref{fig4} where the top frame shows the plot for the STA coupler and the bottom for the original one. We set $\Delta_0=1 \; \text{mm}^{-1}$ and $\Omega_0$ varies between 0 $\text{mm}^{-1}$ and 5 $\text{mm}^{-1}$. We assume that the device length $2L$ varies from 0 mm to 2 mm. The figure shows that the coupler with STA phase mismatch model achieves an efficient and robust light transfer at a shorter device length. For example, for a maximum coupling strength $\Omega_0 = 5 \; \text{mm}^{-1}$, we obtain light intensity $I_2 (L) = 1$ at device length $2L = 0.7 \; \text{mm}$, while the comparable device without STA requires a device length of $2L = 1.3 \text{mm}$. 

\subsection{Robustness against parameter fluctuations}

\begin{figure} [tbhp]
\centering
\includegraphics[width=0.5\textwidth]{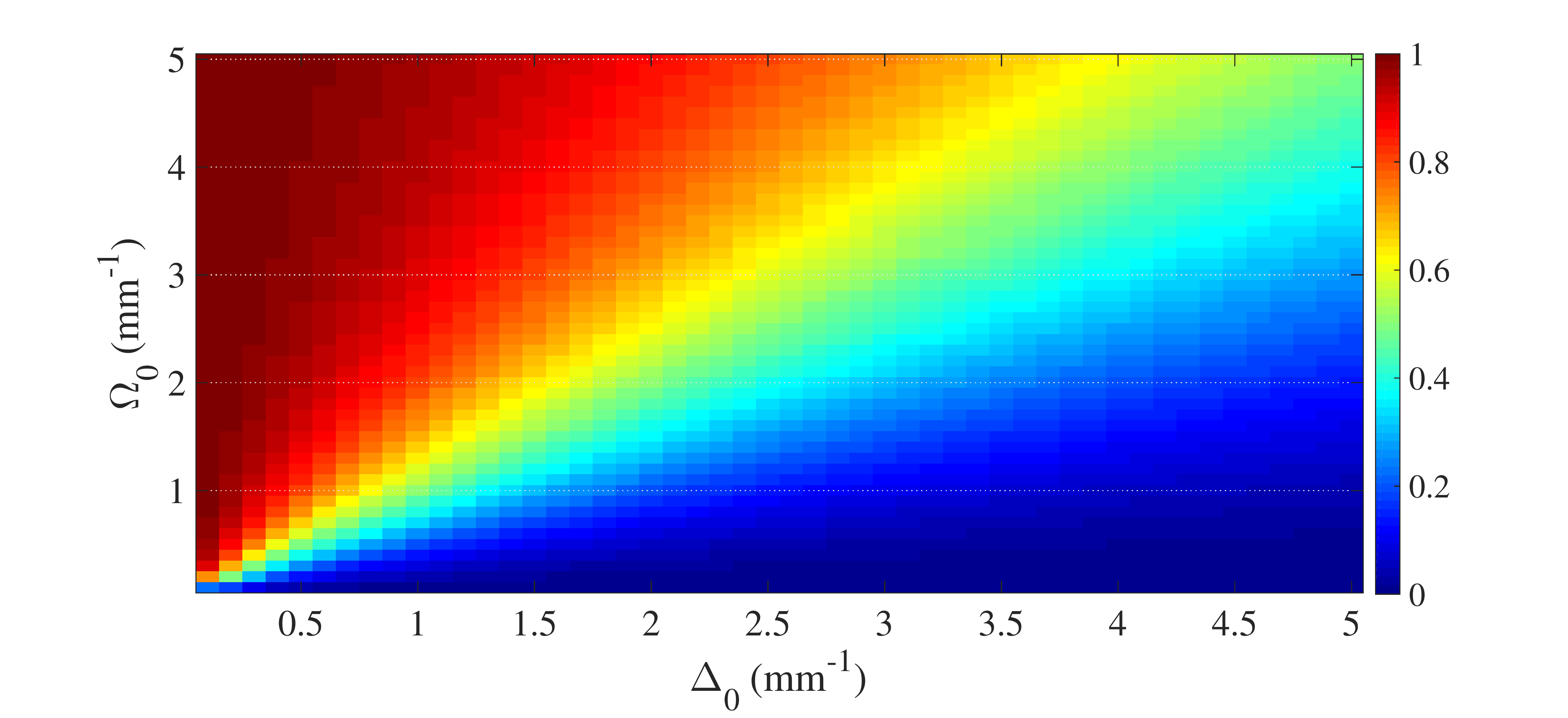}
\includegraphics[width=0.5\textwidth]{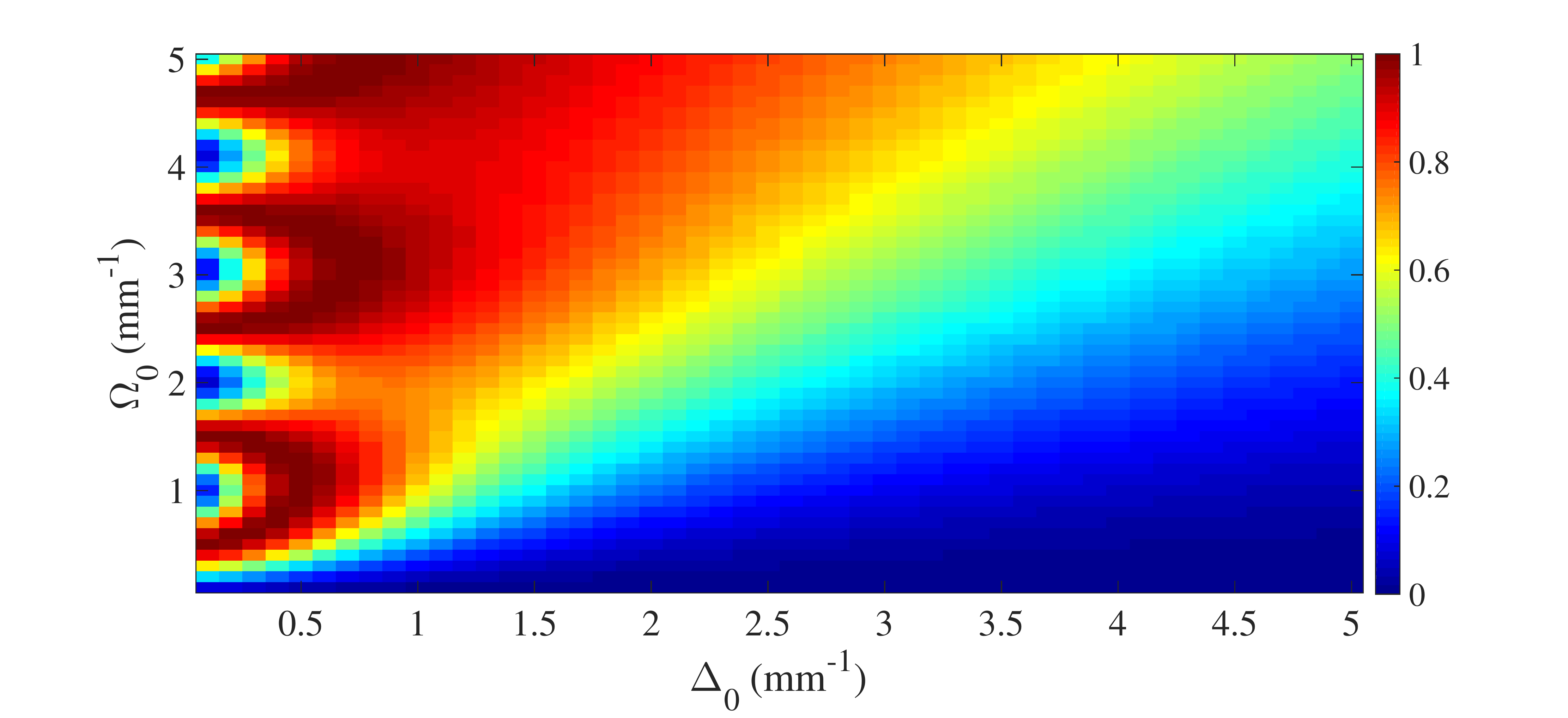}
\caption{(color online) Contour plots of the light intensity transfer $I_2 (L)$. We numerically solve equation \eqref{Schr} for varying maximum coupling strength $\Omega_0$ and phase mismatch $\Delta_0$. The device length is 10 mm. The top frame shows $I_2(L)$ for a coupler with shortcut to adiabaticy, while the bottom frame for a coupler without.}
\label{fig5}
\end{figure}

We continue to show the superiority of the counterdiabatic STA waveguide coupler as compared to the analogous device without STA by examining the light intensity transfer to waveguide 2, $I_2 (L)$, as a function of the maximum coupling strength $\Omega_0$ and the phase mismatch $\Delta_0$. The device length is fixed at $2L = $ 10 mm, while $\Omega_0$ varies from 0 $\text{mm}^{-1}$ to 5 $\text{mm}^{-1}$ and $\Delta_0$ from 0 $\text{mm}^{-1}$ to 5 $\text{mm}^{-1}$, as shown in Fig. \ref{fig5}. We note that the fidelity of the light transfer for the STA coupler is robust against variations in both the maximum coupling and phase mismatch, including around small $\Delta_0$, which is not the case for a light coupler without STA. 

The stability of the light transfer efficiency to changes in the coupling and phase mismatch guarantees the achromatic operation of the proposed coupler. Owing to the fact that different wavelengths of light have different coupling and phase mismatch parameters, Fig. \ref{fig5} clearly shows that these will not affect the fidelity of the light transfer within a moderate wavelength range. 

\section{beam splitter based on shortcut to adiabaticity}
\label{secV}

\begin{figure} [htbp]
\centering
\includegraphics[width=0.3\textwidth]{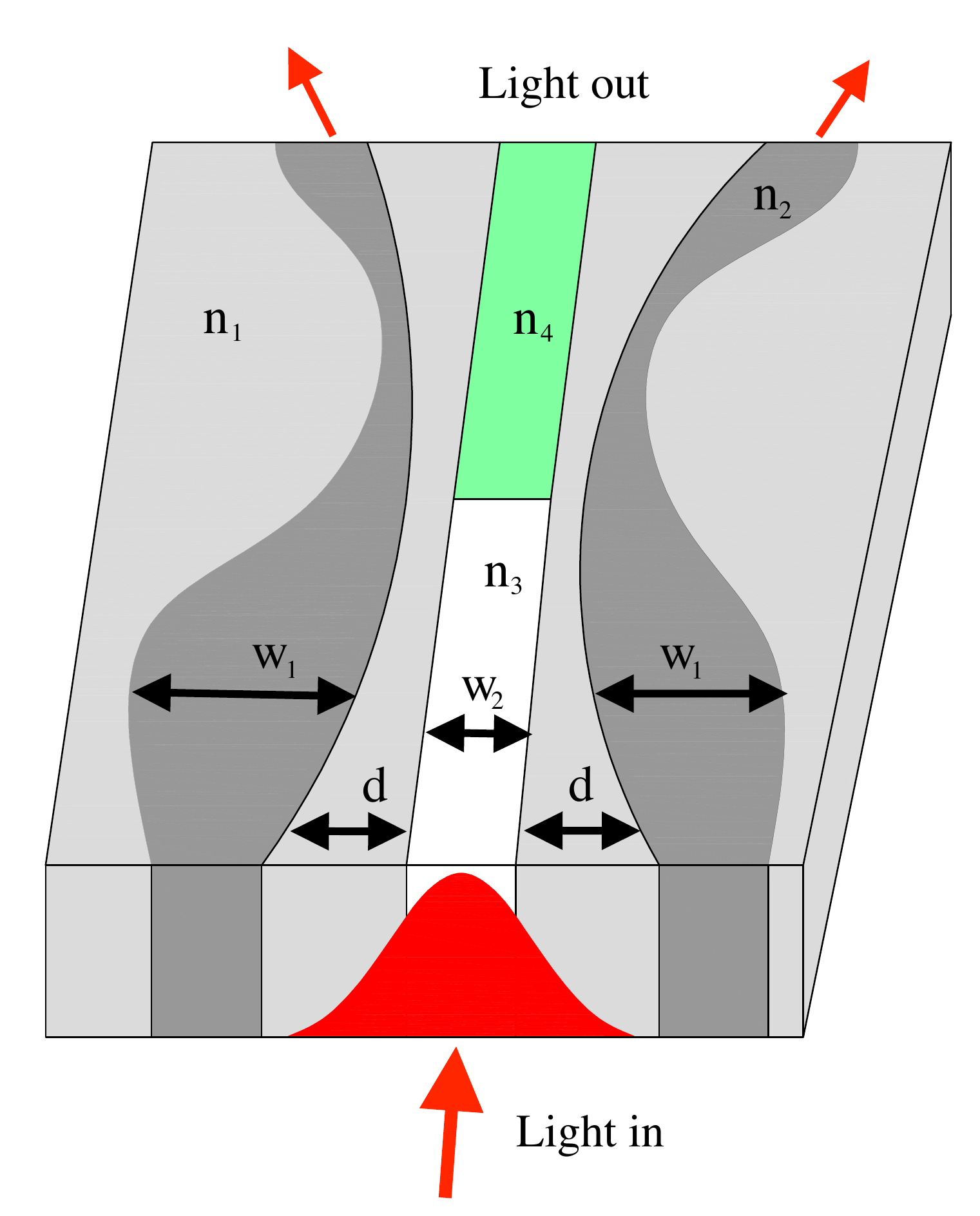}
\caption{(color online) Three evanescently coupled waveguides made of three slabs with refractive indexes $n_2, n_3$ and $n_4$, embedded in a medium with an index of refraction $n_1$. Gaussian-shaped light beam is injected initially in the middle waveguide, which at the end of the evolution, is robustly transferred to the two outer waveguides.}
\label{fig6}
\end{figure} 

In this section, we consider three coupled waveguides as shown in Fig. \ref{fig6}. We assume that the outer waveguides, waveguides 1 and 3, are geometrically symmetric with respect to waveguide 2, that is they are equally coupled to it with $\Omega(z)$. Furthermore, the outer waveguides are assumed to have equal refractive indexes $n_2$, while the refractive index of the middle waveguide changes from $n_3$ to $n_4$ at the maximum of the coupling, $z=0$. We assume that the widths of waveguides 1 and 3 are constant and the width of inner waveguide is a function of $z$. The phase mismatch is defined as $\Delta(z) = \beta_2 (z) - \beta_1 (z)$, where $\beta_1 (z)$ and $\beta_2 (z)$ are the propagation coefficients of waveguides 1 and 2. The light propagation in this waveguide array is described by
\begin{equation}
i\dfrac{d}{dz}\left[
\begin{array}{c}
c_{1}(z) \\
c_{2}(z) \\
c_{3}(z)%
\end{array}%
\right] =%
\begin{bmatrix}
0               & \Omega (z) & 0 \\
\Omega(z) & \Delta (z) & \Omega(z) \\
0               & \Omega (z) & 0 \\
\end{bmatrix}%
\left[
\begin{array}{c}
c_{1}(z) \\
c_{2}(z)\\
c_{3}(z)
\end{array}%
\right] . 
\label{3way}
\end{equation}

These coupled differential equations are analogous to the Schr\"odinger equation describing a three-state quantum system subjected to an external electromagnetic field. Thus, we can introduce a new basis of a dark $c_b (z) = \frac{1}{\sqrt{2}} (c_1(z) + c_3(z) ) $ and a bright state $c_b (z) = \frac{1}{\sqrt{2}} (c_1(z) - c_3(z) )$. Rewriting Eq. \eqref{3way} in the new basis,
\begin{equation}
i\dfrac{d}{dz}\left[
\begin{array}{c}
c_{b}(z) \\
c_{2}(z) \\
c_{d}(z)%
\end{array}%
\right] =%
\begin{bmatrix}
0  & \sqrt{2}\Omega (z) & 0 \\
\sqrt{2}\Omega(z) & \Delta (z) & 0 \\
0   & 0 & 0 \\
\end{bmatrix}%
\left[
\begin{array}{c}
c_{b}(z) \\
c_{2}(z)\\
c_{d}(z)
\end{array}%
\right] ,
\label{dark3}
\end{equation}
we can easily see that the dark state $c_d(z)$ is decoupled, and the three-state problem is reduced to a two-state one involving states $c_b(z)$ and $c_2(z)$ only. 

This set of differential equations is the same as the one for two coupled waveguides. Therefore, if we use the same parameters with shortcut to adiabaticity $\Omega(z)$ and $\Delta(z)$ from Eq. \eqref{shortcut}, we can realize a robust and fast complete state transfer from $c_2 (z)$ to $c_b (z)$. Mapping to the three coupled waveguides system, if we assume that light is initially input in the middle waveguide, then the final output light will be in state $c_b(z)$, which realizes an equal intensity superposition between the two outer waveguides. Note that by design, this device will have the same advantages as the two-waveguide coupler, these are {\it i)} a shorter device length; {\it ii)} achromaticity; and {\it iii)} robustness to parameter fluctuations including around small values for the phase mismatch. 

\section{Conclusions}  
\label{secVI}

We demonstrated a novel device for complete achromatic optical switching between evanescently coupled waveguides. Utilizing counterdiabatic STA-obtained changes to the coupling and the phase mismatch, we show that the proposed device realizes a complete and robust achromatic light switching on a shorter length scale as compared to previous designs. We note that the required parameter changes obey the coupling strength inequality, $|\Omega_a (z)| \le |\Omega (z) | \le |\Omega_0 (z)|$. Finally, we showed that a similar waveguide coupler with STA can be used for realizing an equal superposition beam splitter in a system of three coupled waveguides. 

\section{Acknowledgements}

W. H. and L. K. A. are partly supported by Singapore ASTAR AME IRG A1783c0011 and U.S. Air Force Office of Scientific Research (AFOSR) through the Asian Office of Aerospace Research and Development (AOARD) under Grant No. FA2386-17-1-4020. 

E. K. acknowledges financial support from the European Union's Horizon 2020 research and innovation programme under the Marie Sk\l odowska-Curie grant agreement No 705256 --- COPQE.


\end{document}